\documentclass[fleqn,10pt]{wlscirep}
\usepackage{multicol}
\usepackage{multirow}
\usepackage{textcomp}
\usepackage{graphics}

\title{Simulating and Optimising Quantum Thermometry Using Single Photons}

\author[1,*]{W.K. Tham}
\author[1]{H. Ferretti}
\author[1]{A.V. Sadashivan}
\author[1,2]{A.M. Steinberg}
\affil[1]{Centre for Quantum Information \& Quantum Control and Institute for Optical Sciences, Department of Physics, University of Toronto, 60 St. George St, Toronto, Ontario, Canada, M5S 1A7}
\affil[2]{Canadian Institute For Advanced Research, 180 Dundas St. W., Toronto, Ontario, Canada, M5G 1Z8}

\affil[*]{wtham@physics.utoronto.ca}

\begin{abstract}
A classical thermometer typically works by exchanging energy with the system being measured until it comes to equilibrium, at which point the readout is related to the final energy state of the thermometer. A recent paper noted that different-temperature baths lead not only to different equilibrium states but also to different equilibration \emph{rates}. In some cases this means that temperature discrimination is better achieved by comparing the rates than the asymptotic states -- and should therefore be carried out at finite times rather than once equilibration is essentially complete. The theory work also noted that for a \emph{quantum} thermometer, the difference between the relaxation rates for populations and coherences means that for intermediate time regimes (before full equilibration but after some characteristic time that depends on the temperatures of the baths), optimal discrimination is achieved not by probing energy only but by using quantum coherence as well. In this work, we study these effects experimentally. Implementing a recent proposal for efficiently emulating an arbitrary quantum channel, we use the quantum polarisation state of individual photons as models of ``single-qubit thermometers'' which evolve for a certain time in contact with a thermal bath. We investigate the optimal thermometer states for temperature discrimination, and the optimal interaction times, confirming that there is a broad regime where quantum coherence provides a significant improvement. We also discuss the more practical question of thermometers composed of a finite number of spins/qubits (greater than one), and characterize the performance of an adaptive protocol for making optimal use of all the qubits.
\end{abstract}
\begin{document}

\flushbottom
\maketitle
\thispagestyle{empty}

\section*{Introduction}

A multitude of measurement and metrology tasks have been shown to benefit, sometimes dramatically, from the substitution of certain classical resources with their quantum counterpart\cite{Lloyd2004SQL,Lloyd2011QuantumMetrologyAdvances}. Examples of advances in quantum metrology range from many-fold increase in sensitivity to phase in interferometry or polarization in polarimetry through the use of non-classical light\cite{aasi2013enhanced,lee2002quantum,AE2004N00NSuperRes,walther2004broglie,nagata2007beating,tsang2009quantum,LeeDylan2014N00NCentroid,PhysRevLett.107.083603} to highly sensitive magnetometry\cite{jaklevic1964SQUIDs,auzinsh2004can,jones2009magnetic}. Much in keeping with the spirit of the field, a recent analysis\cite{rudolphThermometry2015} has considered that most primitive of metrological tasks - thermometry, or simply telling cold from hot - and found that a coherent measurement scheme can enhance thermometry beyond the traditional approach of allowing a measurement device to fully thermalize and equilibrate with the thermal bath being measured.

Our measurement device in this case is a qubit or a quantum system consisting of two levels typically called the ground (denoted $\left|0\right>$) and excited ($\left|1\right>$) states. Furthermore, since a single spin cannot provide more than one bit of information\cite{holevo1973bounds}, we follow \cite{rudolphThermometry2015} in considering the simpler task of distinguishing between just \emph{two} (instead of a continuum of) thermal baths at two different temperatures. And whereas a traditional thermometry approach prescribes letting the thermometer fully thermalise to the given bath by waiting for a long time, whereupon some physical quantity that bears a known correlation with the temperature is measured, we allow our qubit to interact for only a finite duration before it is subjected to some measurement. The qubit interacts with the heatbath by absorbing or emitting photons from/into it. In standard treatments of spontaneous emission\cite{Carmichael2009OpenSystems,mandel1995optical} this photon exchange process is often described with parameters $\tau_{1}$, the decay time for the excited state population, and $\tau_{2}$, the decay time for coherences between $\left|0\right>$ and $\left|1\right>$. It is known that $2\tau_{1}\geq\tau_{2}$. When there are no additional dephasing mechanisms, the equality holds so that the coherence damps away exactly half as quickly as the excited state population. In the case of thermalisation via energy exchange with a bosonic bath, as treated in\cite{rudolphThermometry2015}, the collision rate grows with occupation number, and hence with temperature. $\tau_{1}$ and $\tau_{2}$ are therefore shorter for higher-temperature baths. This difference means that it is generally advantageous to wait for a finite interaction time (on the order of the thermalisation times) in order to better distinguish the two processes, rather than allowing the qubit to fully thermalise with the bath. In particular, for high temperatures, the asymptotic populations are very similar, while the thermalisation rates may be quite different. Since coherences decay twice as slowly as populations, the time of optimal distinguishability occurs roughly twice as late for initially coherent states (which decay principally at $\tau_{2}$) as for initially incoherent states (which decay at $\tau_{1}$). As a consequence, one can show that beyond a certain critical time, an initial state with some coherence always makes for a more sensitive thermometer. Similar results are expected to hold in any case where thermalisation occurs faster with higher-temperature baths (for instance, via collisions with Maxwell-Boltzmann distributed gas molecules), but not if the thermalisation rate were fixed and temperature-independent, as in commonly used phenomenological models of thermal conductance. (For the interested reader, we give a mathematical argument in the Supplementary, as to why these observations are true). This work therefore aims to experimentally demonstrate this metrological advantage, along with an extension to the more practically relevant case where one is not restricted to the use of a single qubit.

\begin{figure}
    \noindent \begin{centering}
    \includegraphics[width=18cm]{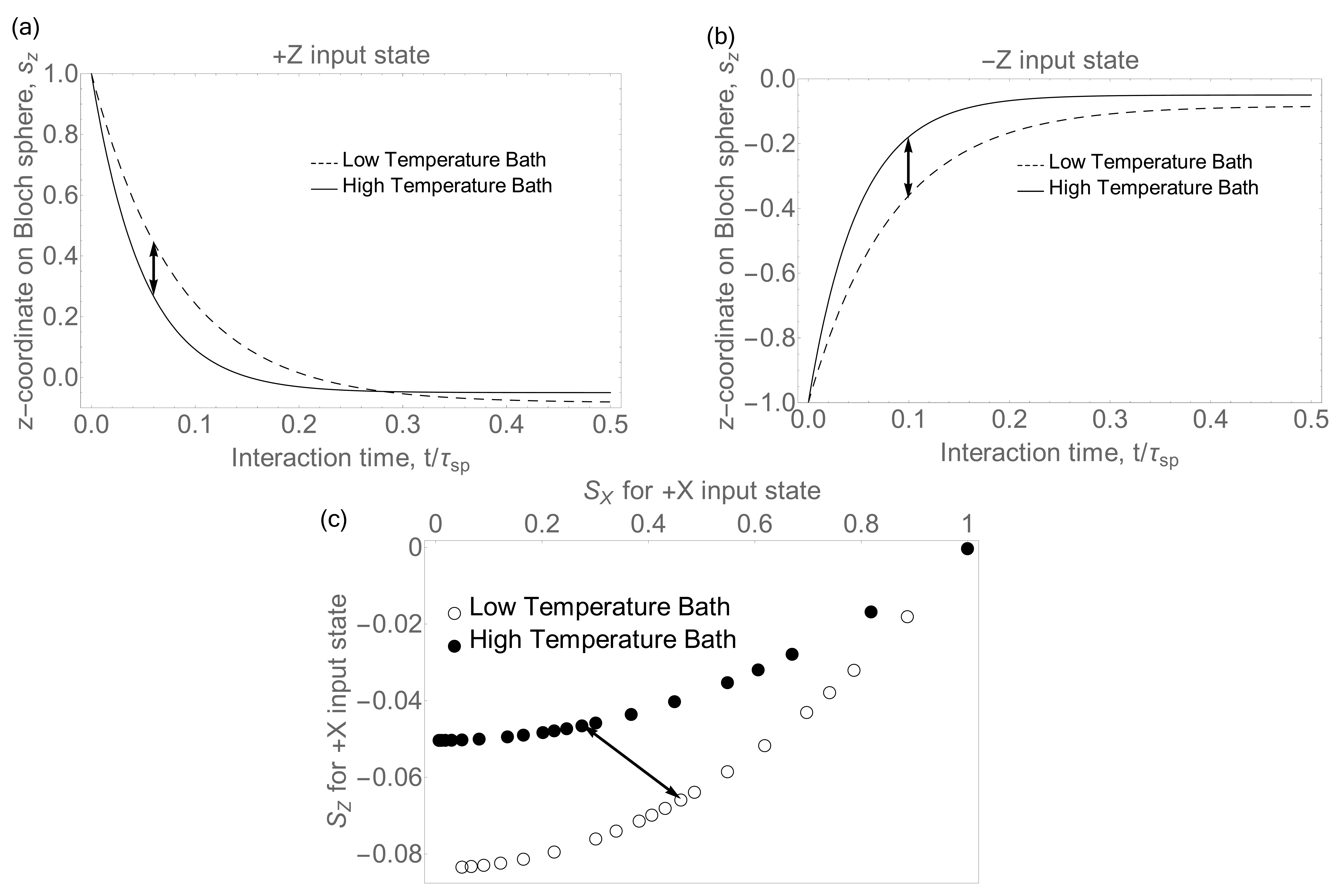}
    \par\end{centering}

    \caption{\textbf{Bloch vector components vs interaction time.} Theoretically computed components of the Bloch vector after thermalizing for $t$ seconds,\textbf{ (a)} given a $+Z$ input state, \textbf{(b)} $-Z$ input state, and \textbf{(c)} $+X$ input state. For this latter case, both $s_{z}$ and $s_{x}$ are shown. At $t=0$ the state begins at the rightmost point of the bottom plot. Each subsequent timestep is shown as a pair of points, one each for high and low temperatures respectively. Arrows indicate where the greatest separation occurs.\label{fig:Bloch-vector-components}}
\end{figure}

In the Bloch sphere representation, every single qubit state corresponds uniquely to a 3-vector on or within the unit sphere. Conventionally, the excited state is represented as $+Z$ and the ground state as $-Z$. Maximally coherent states lie on the $xy$-plane, usually with $\left(\left|0\right>+\left|1\right>\right)/\sqrt{2}$ at $+X$. The thermalisation of the qubit in this picture can be thought of as a trajectory from its initial Bloch vector (usually a unit vector on the surface of the unit sphere if the initially prepared state is pure) to its final point on the $z$-axis. To aid the reader in visualising this, figure \ref{fig:Bloch-vector-components} shows trajectories of a qubit initialised in $+Z$, $-Z$, and $+X$. For the $\pm Z$ initial states, the evolution of the qubit is strictly along the $z$-axis, so only the $z$ component (denoted $s_{z}$) is shown. When the qubit is initialised to $+X$, however, both the coherence (the $x$-component, labeled $s_{x}$) and the excited state population (the $z$-component) relax with time so both are shown in figure \ref{fig:Bloch-vector-components}(c) over a range of discrete times.

The yardstick by which we will characterise the performance of our qubit thermometer is the probability with which we mis-identify our bath, either by mistaking a hot bath for a cold one or vice versa. Since a thermalised qubit is almost always in a mixed state (a statistical mixture of pure states), this error probability or $p_{e}$ never vanishes. For example, a fully thermalized qubit at temperature $T=\infty$ is in a state that is an equal statistical mixture between $\left|0\right>$ and $\left|1\right>$ whereas at $T=0$ it is in the ground or $\left|0\right>$ state. Suppose we now identify our heatbath as the hot one if and only if a measurement on the qubit finds it in the excited state. Although we will never misidentify the $T=0$ bath, the $T=\infty$ bath yields the excited state with $50\%$ probability so we stand to misidentify it half the time! Assuming that a given bath is chosen from $T=\infty$ and $T=0$ with equal likelihood, our overall error probability, $p_{e}$, is 1/4. Tasks such as the one just described are aptly called state discrimination. Conveniently, $p_{e}$ in state discrimination is well-known to be related to $\vec{r}_{1}$ and $\vec{r}_{2}$, Bloch vectors corresponding to the states being discriminated, as follows\cite{Fuchs1996Thesis,NielsenChuang2010}:
\begin{equation}
    1-p_{e}\leq\frac{1}{2}\left(1+\frac{1}{2}\left|\vec{r}_{1}-\vec{r}_{2}\right|\right)\label{eq:EuclideanDistancePe}
\end{equation}
where the norm is to be understood as the usual Euclidean/Cartesian distance between vectors $\vec{r}_{1}$ and $\vec{r}_{2}$. Returning to our example with $T_{hot}=\infty$ and $T_{cold}=0$, the completely mixed state corresponds to $\vec{r}_{1}=\left\langle 0,0,0\right\rangle $ and the ground state to $\vec{r}_{2}=\left\langle 0,0,-1\right\rangle $, so $\left|\vec{r}_{1}-\vec{r}_{2}\right|=1$ implying once more that $p_{e}\geq1/4$. It is important to stress that while a larger Euclidean distance between Bloch vectors implies that a lower error probability is achievable in principle, actually saturating the inequality to achieve the lowest possible error requires that we select the correct basis during measurement. In our example above, measuring if the qubit is in the state $\left(\left|0\right>+\left|1\right>\right)/\sqrt{2}$ would have yielded a ``yes'' answer with probability $50\%$ for \emph{both} $T=0$ and $T=\infty$, giving us no information at all about the bath!

\section*{Experimental Design}
\subsection*{Emulating thermalisation with photons}

In designing a tabletop experiment to test the use of coherence and adaptivity in qubit thermometry, we have opted to optically simulate the heatbaths instead of subjecting our qubits to actual thermalization. To do so, we must first understand how thermalisation affects qubits. In the absence of extraneous damping processes (e.g. mechanical collision), a qubit that interacts with a thermal reservoir by photon exchange alone can be treated as a system that emits or absorbs a photon into/from the reservoir with some probability. Such a process is well-modeled by a generalised amplitude damping (GAD) channel, which is defined in the standard operator-sum representation as follows\cite{NielsenChuang2010}:
\begin{equation}
    \rho_{thermalized}=p\left(K_{11}\rho_{initial}K_{11}^{\dagger}+K_{12}\rho_{initial}K_{12}^{\dagger}\right)+\left(1-p\right)\left(K_{21}\rho_{initial}K_{21}^{\dagger}+K_{22}\rho_{initial}K_{22}^{\dagger}\right)\label{eq:Map}
\end{equation}

\begin{multicols}{2}
    \setlength{\parskip}{0mm}
    \begin{eqnarray*}
    K_{11} & = & \left[\begin{array}{cc}
    1 & 0\\
    0 & \sqrt{1-\gamma}
    \end{array}\right]\\
    K_{12} & = & \left[\begin{array}{cc}
    0 & \sqrt{\gamma}\\
    0 & 0
    \end{array}\right]
    \end{eqnarray*}

    \begin{eqnarray}
    K_{21} & = & \left[\begin{array}{cc}
    \sqrt{1-\gamma} & 0\\
    0 & 1
    \end{array}\right]\nonumber \\
    K_{22} & = & \left[\begin{array}{cc}
    0 & 0\\
    \sqrt{\gamma} & 0
    \end{array}\right]\label{eq:Krauses}
    \end{eqnarray}
\end{multicols}Equivalently we can write the above process in terms of its action on the Bloch vector:
\begin{equation}
    \vec{v}_{thermalized}=\left[\begin{array}{ccc}
    \sqrt{1-\gamma} & 0 & 0\\
    0 & \sqrt{1-\gamma} & 0\\
    0 & 0 & \left(1-\gamma\right)
    \end{array}\right]\vec{v}_{initial}+\left[\begin{array}{c}
    0\\
    0\\
    \left(2p-1\right)\gamma
    \end{array}\right]\label{eq:BlochMap}
\end{equation}
In words, the Kraus operators describe two physical sub-processes: $K_{11}$ and $K_{12}$ jointly describe a sub-process in which the qubit in $\left|0\right>$ \emph{absorbs} a photon from the reservoir with probability $\gamma$ thereby transitioning to $\left|1\right>$. $K_{21}$ and $K_{22}$ describe the opposite sub-process in which a photon is \emph{emitted} into the reservoir again with probability $\gamma$. A thermalising qubit is merely one which undergoes the first sub-process (absorption) with probability $p$ and the second (emission) with probability $1-p$. The probability $p$ in turn is determined by the bath temperature. To see this, suppose our qubit states $\left|0\right>$ and $\left|1\right>$ have energies $E_{0}$ and $E_{1}$ respectively. After fully thermalising, we expect our qubit, which is now in a mixture of $p\left|1\right>$ and $\left(1-p\right)\left|0\right>$, to obey thermal statistics. We expect:
\begin{eqnarray*}
p & = & \frac{\exp\left(-E_{1}/k_{B}T\right)}{\exp\left(-E_{0}/k_{B}T\right)+\exp\left(-E_{1}/k_{B}T\right)}\\
\implies\ln\frac{1-p}{p} & = & \frac{E_{1}-E_{0}}{k_{B}T}=\frac{\hbar\omega}{k_{B}T}
\end{eqnarray*}
so $p\to0$ as $T\to0$ whereas $p\to1/2$ as $T\to\infty$. It is also well-known that for a bosonic thermal reservoir, the Planck distribution implies an average occupation number, $\overline{N}=\left[\exp\left(\hbar\omega/k_{B}T\right)-1\right]^{-1}$. Thus, in terms of $\overline{N}$, we can write $p$ more compactly as $p=\overline{N}/\left(1+2\overline{N}\right)$.

The damping parameter $\gamma$ is related to the interaction time with the bath, $t$, and temperature, $T$, as follows: $\gamma=1-\exp\left(-t/\xi\tau_{sp}\right)$, where $\tau_{sp}$ is a timescale characteristic of the coupling between qubit and bath. Here,
\begin{eqnarray*}
    \xi & = & \tanh\left(\hbar\omega/2k_{B}T\right)=\frac{1}{1+2\overline{N}}\label{eq:timeConstant}
\end{eqnarray*}
is a unitless quantity that encodes the bath temperature. Note from equation \ref{eq:timeConstant} that $\xi$, and therefore the relaxation rate of the qubit's excited state population and coherence, is temperature dependent. This, coupled with the fact that the difference of two exponential functions with different exponents is \emph{not} monotonic, means that the Euclidean distance between resultant states for the hot vs cold baths is larger (and our thermometer more sensitive) when the interaction time $t$ is finite (partially thermalised) as opposed to infinite (fully thermalised).

Since in our case we are merely emulating thermalisation, $\omega$ which defines the mode through which the qubit is coupled to the bath, is ill-defined. We shall therefore specify temperature in terms of $\overline{N}$ and $\hbar\omega/k_{B}$. Likewise, we do not have an intrinsic timescale by which to specify $\tau_{sp}$. All times $t$ will be specified in units of $\tau_{sp}$. As we'll see, encoding the qubit in two orthogonal polarisations of a photon (call them $\left|H\right>$ and $\left|V\right>$) makes it possible to implement this compactly in a tabletop experiment.

\subsection*{Implementation}

\begin{figure}
    \noindent \begin{centering}
    \includegraphics[width=18cm]{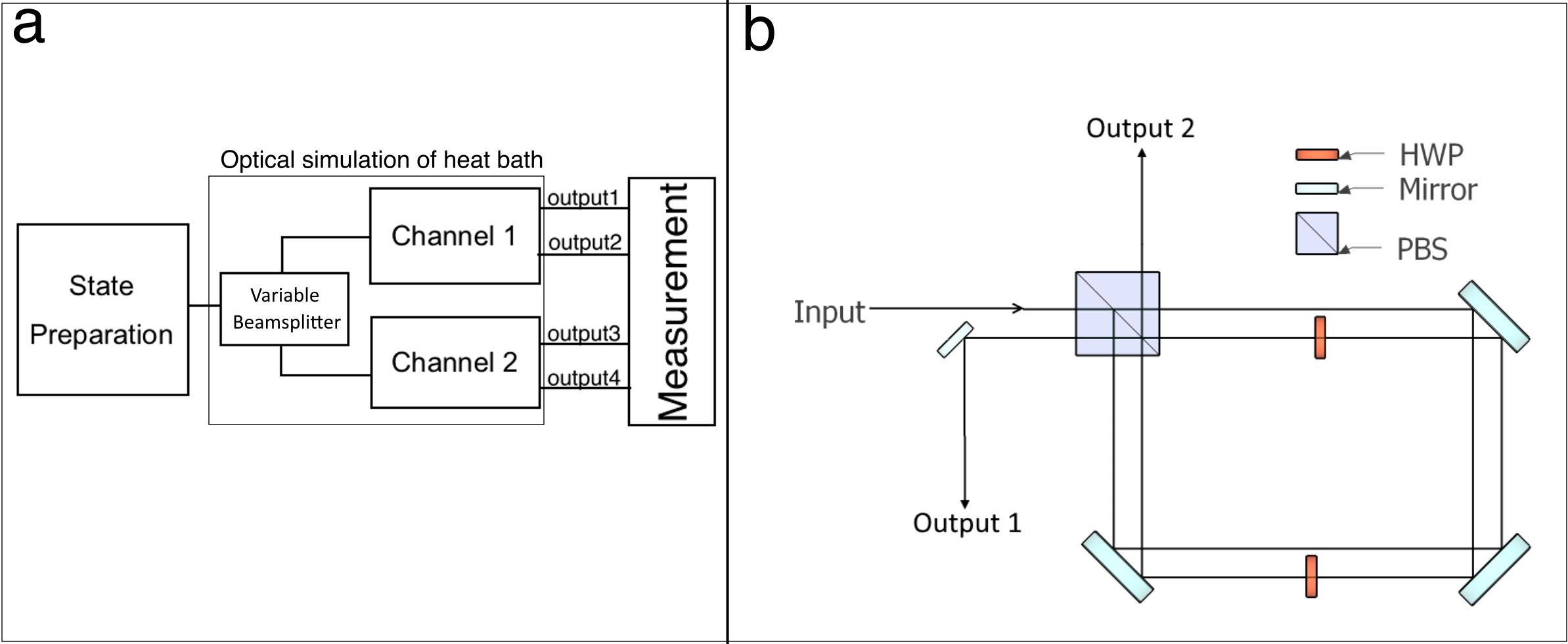}
    \par\end{centering}

    \caption{\textbf{Experimental Scheme. a)} Block schematic of the experimental setup for channel emulation.\textbf{ b)} Drawing of an optical implementation of the ``channels'' block. Two similarly constructed copies of the channel driven by a splitter that switches between them implements a simulation of a thermal bath. The ``switching'' is done via a variable beamsplitter (VBS), whose coefficient of reflection/transmission can be modulated.\label{fig:SagnacSchematic}}
\end{figure}

We use a type-II spontaneous parametric down-conversion (SPDC) setup with a $2mm$ BBO crystal pumped by a $50mW$ continuous wave $405nm$ diode, phase-matched with an opening angle of 3\textdegree, and filtered with 10nm bandpass filters. One photon of the SPDC pair is sent directly to a single-photon counting module (SPCM) to act as herald whereas the other is used as a $810nm$ light source for our experiment. Typical total coincidence rates (taking into account all losses in the experiment) are $\approx4000$ counts per second. The first stage of the experiment immediately following the SPDC source is a polariser $\to$ quarter-wave plate (QWP) $\to$ half-wave plate (HWP) sequence, which collectively comprises the ``state preparation'' block in figure \ref{fig:SagnacSchematic}.

To simulate thermalization, we make use of an optical circuit having the capacity to simulate any valid single-qubit quantum channel (precisely, completely positive trace-preserving or CPTP maps). The design of this circuit was inspired by Sanders et al's theoretical results\cite{BarrySanders2013Channels}, which were in turn based on mathematical work by Ruskai et al\cite{Ruskai2002analysis}. A similar optical circuit has already been shown to accurately simulate a wide variety of single-qubit quantum channels\cite{lu2015universal}. Our design, illustrated in figure \ref{fig:SagnacSchematic} consists of a variable beamsplitter (VBS), followed by two subsequent interferometers (labeled channels 1 and 2 respectively in figure \ref{fig:SagnacSchematic}). The optical design of each channel is shown in inset (b) of figure \ref{fig:SagnacSchematic}. Light that is incident on each channel is split at a polarising beamsplitter (PBS) so that each of two orthogonal polarisations travels along spatially separated counter-propagating paths. Each path contains a HWP, allowing the polarisations to be rotated independently before being recombined at the PBS. It is easy to see how this intra-interferometer rotation allows us to emulate a damping channel. Suppose we leave the $\left|H\right>$ polarisation within the interferometer unrotated but leave the HWP within the $\left|V\right>$ path at 45\textdegree{} so that $\left|V\right>\to\left|H\right>$. Such a setting guarantees that the output of the channel is \emph{always} $\left|H\right>$ regardless of input polarisation - hence a full damping channel with $\gamma=1$. Less extreme settings realise the full range of damping channels. More formally, it is easy to show that each channel is well described by $\rho_{output}=K_{1}\rho_{in}K_{1}^{\dagger}+K_{2}\rho_{in}K_{2}^{\dagger}$ with:
\begin{multicols}{2}
    \setlength{\parskip}{0mm}
    \begin{eqnarray*}
    K_{1} & = & \left[\begin{array}{cc}
    \cos2\theta_{H} & 0\\
    0 & \cos2\theta_{V}
    \end{array}\right]
    \end{eqnarray*}

    \[
    K_{2}=\left[\begin{array}{cc}
    0 & \sin2\theta_{V}\\
    \sin2\theta_{H} & 0
    \end{array}\right]
    \]
\end{multicols}
Here $\theta_{H}$ and $\theta_{V}$ are angles of the fast axis of the half-wave plates in channels 1 and 2, acting on the horizontal and vertical paths respectively. Setting $\theta_{H}=0$ and $\sin2\theta_{V}=\sqrt{\gamma}$ implements $K_{11}$ and $K_{12}$ in equation \ref{eq:Krauses}, whereas setting $\theta_{V}=0$ and $\sin2\theta_{H}=\sqrt{\gamma}$ implements $K_{21}$ and $K_{22}$.

Although the inner workings of the VBS are not shown, it is identical in design to the channels except for the fact that there is a single HWP (instead of two independent ones) acting on both counter-propagating paths. This restricts the action of the VBS to a \emph{fixed} unitary that can be subsequently undone for all input states with a simple HWP placed outside the VBS. Notice that if we now set the HWP in the VBS to $\cos^{2}2\theta_{VBS}=p$, both channels work in tandem to fully implement the map in equation \ref{eq:Map}. Thus, we have a fully tunable means of emulating thermalisation with a polarisation qubit.

Finally, the measurement block consists of a QWP $\to$ HWP $\to$ PBS sequence. The four output states (two from each channel, one from each Kraus operator) are mixed and sent through this measurement sequence and then onto an APD and coincidence counter. In practice, losses and imperfections in optical components means that mixing the four output states \emph{before} the measurement sequence is impractical since it precludes the possibility of compensating with post-processing. Instead, we opted to send each one to two APDs after the measurement sequence (they are further split at the PBS into $\left|H\right>$ and $\left|V\right>$) and then only tracing over them in post-processing. Because the number of settings for state preparation, measurement, and channel selection is potentially vast, the HWPs and QWPs are mounted on motorised rotation mounts (Thorlabs PRM1/MZ8) driven by DC servo controllers (Thorlabs KPRM1E/M) where necessary. In addition to automating time consuming parts of the experiment, they provide the added benefit of more precise angular settings ($\pm0.2\text{\textdegree}$ as opposed to roughly $\pm0.5\text{\textdegree}$ when done manually).

\section*{Experimental Data}

\begin{figure}
    \noindent \begin{centering}
    \includegraphics[width=19cm]{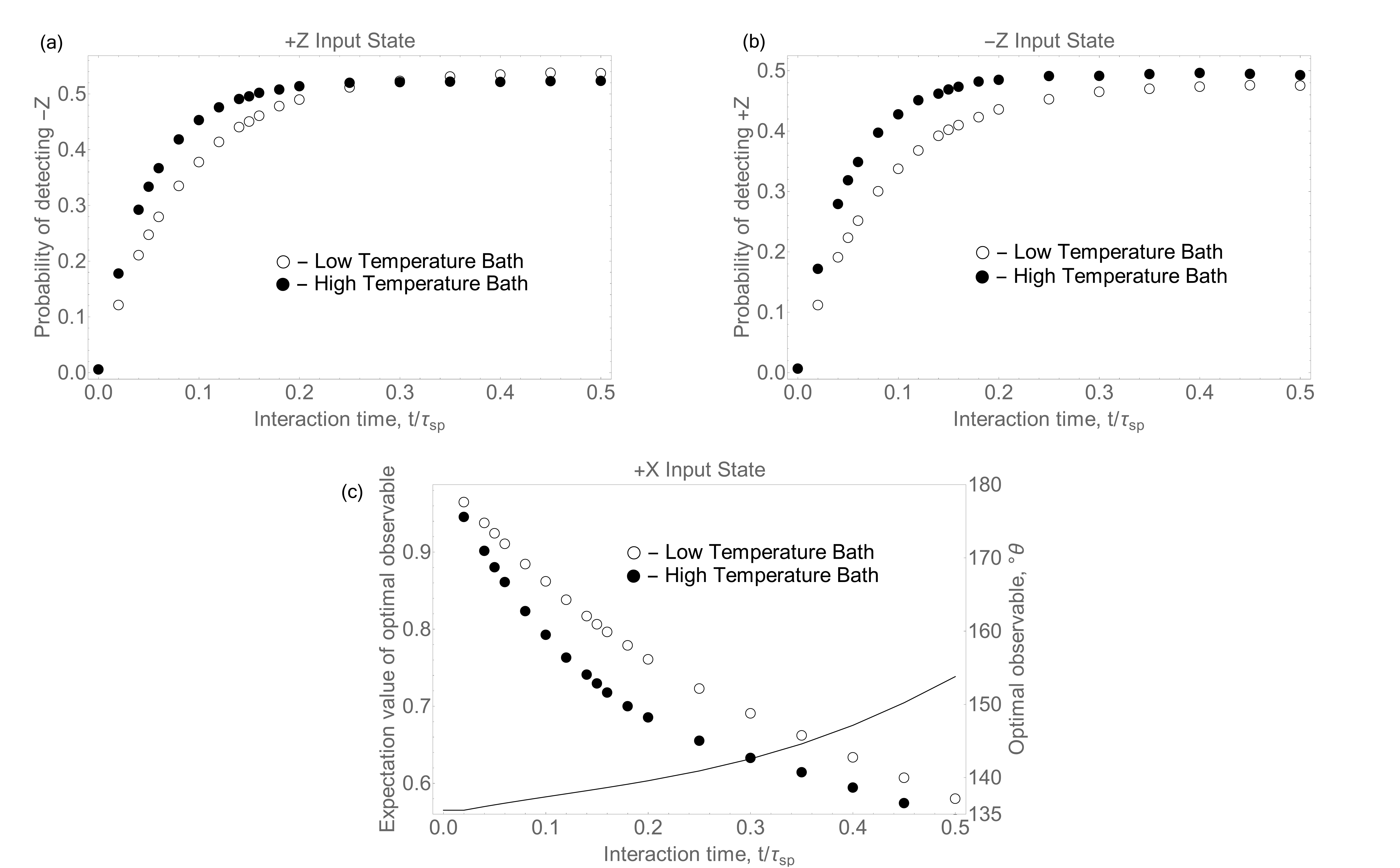}
    \par\end{centering}

    \caption{\textbf{Plots of inferred detection probabilities. }For each input state and channel setting, we computed and performed the optimal measurement for discrimination between the final states from each heat bath. \textbf{(a)} Detection probability of the $-Z$ observable (i.e. $\left|H\right>\left<H\right|$) when input state is $+Z$ (or $\left|V\right>$). \textbf{(b)} Detection probability of $+Z$ observable (i.e. $\left|V\right>\left<V\right|$) when input state is $-Z$ (or $\left|H\right>$). \textbf{(c)} Detection probability for some optimal measurement given input $+X$ (or $\left|D\right>$). Although for $\pm Z$ input states the optimal observable is fixed, in the $+X$ case it varies with interaction time. In this plot, the optimal observable is parametrised as: $\left|\theta\right>\left<\theta\right|$ where $\left|\theta\right>=\cos\theta\left|H\right>+\sin\theta\left|V\right>$. For reference, the relaxation times for the hot bath are $\tau_{2}=2\tau_{1}=1/20=0.05\tau_{sp}$. For the cold bath, they are $\tau_{2}=2\tau_{1}=1/12\approx0.083\tau_{sp}$.
    \label{fig:PlotsRaw}}
\end{figure}

To ensure that our channels are emulating the desired thermalisation process, we characterise it via full process tomography\cite{mohseni2008quantum,kosut2004optimal,jevzek2003quantum,sacchi2001maximum}. Throughout stretches of data-taking, we re-characterise periodically (approximately every 30 minutes, the minimum time within which visibility of interferometers are likely to have dropped appreciably) and realign optics as necessary. We proceeded to prepare states $-Z$ (or $\left|H\right>$), $+Z$ (or $\left|V\right>$), and $+X$ (or $\left|D\right>=\left(\left|H\right>+\left|V\right>\right)/\sqrt{2}$). We emulate the heat baths specified in \cite{rudolphThermometry2015}, with $\xi_{cold}=1/\left(1+2\overline{N}\right)=1/12$ and $\xi_{hot}=1/20$. These correspond to temperatures of $\approx6\hbar\omega/k_{B}$ and $10\hbar\omega/k_{B}$ respectively. We further set the channels to emulate interaction times ranging from $t=0\tau_{sp}$ to $t=0.4\tau_{sp}$, where asymptotics have yet to dominate and dynamics are non-trivial. The time steps are sampled unevenly because: a) at large $t$'s, we were limited by the precision of our motorised rotation stage ($\sim0.2\text{\textdegree}$) whereas b) at small $t$'s, we restricted ourselves to $\Delta t\geq0.2\tau_{sp}$ in order to maintain a reasonable sampling stepsize. For each channel setting, we chose to measure along a basis prescribed by static state discrimination strategies to be optimal (i.e. select a projector $M$ s.t. equation \ref{eq:staticPe} is minimized, discussed in next section). We counted photons for $10$ seconds per measurement, which yielded measurement ``shots'' that consist of $\sim40,000$ photons apiece. Although the thermometry scheme discussed above is intended for single qubits, we have opted to use a bright source and long count durations in order to infer $\mbox{Tr}\left(\rho M\right)$, the probability of a successful outcome of the projector, $M$, in our chosen measurement basis. Since we expect the number of coincidences to be binomially distributed, this inferred detection probability tends to the true single-photon probability with diminishing uncertainty as the total photon number becomes large. Figures \ref{fig:PlotsRaw} show this inferred probability. Note that each point in the plots represents an average over many sets of data taken under identical experimental conditions (9 sets for the $+Z$ case, $10$ sets for $+X$, and $4$ for $-Z$, where a hardware issue forced us to discard $5$ sets of data; discarded sets are reported in the Supplementary, for completeness).

\begin{figure}
    \noindent \begin{centering}
    \includegraphics[width=16cm]{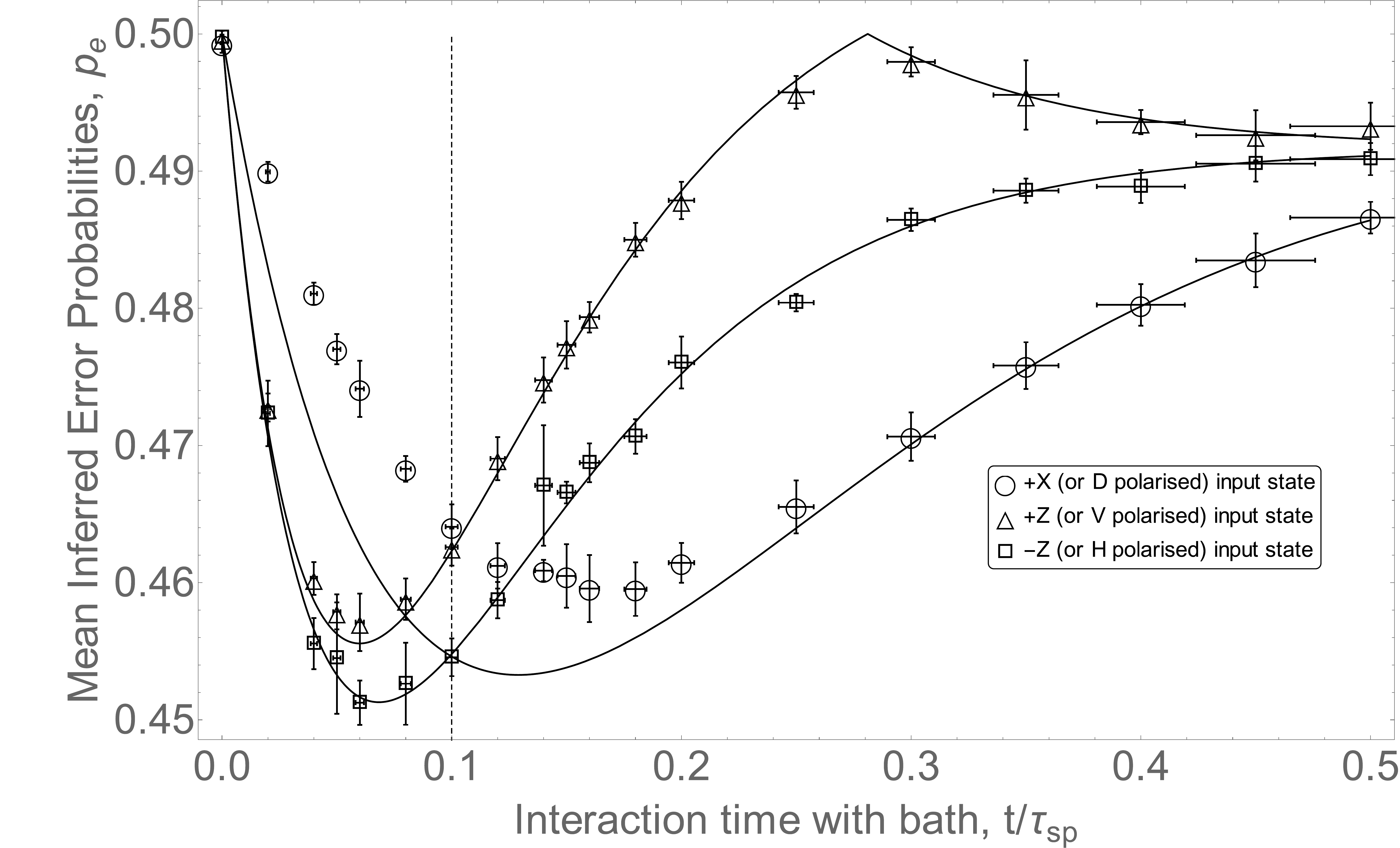}
    \par\end{centering}

    \caption{\textbf{Inferred Error Probabilities} in discriminating between final states from two heat baths. Solid lines are theory curves. Points are experimentally inferred error probabilities. Vertical error bars indicate standard deviation or spread range over multiple repetitions for a given channel setting and input state. Horizontal error bars indicate uncertainty in channel setting due to finite precision of motorised rotational mounts. Relaxation times for the hot bath are $\tau_{2}=2\tau_{1}=1/20=0.05\tau_{sp}$. For the cold bath, they are $\tau_{2}=2\tau_{1}=1/12\approx0.083\tau_{sp}$.\label{fig:InferredEuclidean}}
\end{figure}

In order to compare with theory, the discrimination error probability $p_{e}$ can be computed from the above detection probabilities as:
\begin{equation}
    p_{e}=\min\left\{ 0.5\mbox{Tr}\left(\rho_{1}M\right)+0.5\left(1-\mbox{Tr}\left(\rho_{2}M\right)\right),0.5\left(1-\mbox{Tr}\left(\rho_{1}M\right)\right)+0.5\mbox{Tr}\left(\rho_{2}M\right)\right\} \label{eq:InferredPe}
\end{equation}
where $M=\left|\theta\right>\left<\theta\right|$, $\left|\theta\right>=\cos\theta\left|H\right>+\sin\theta\left|V\right>$, and $\mbox{Tr}\left(\rho M\right)$ are detection probabilities shown in figure \ref{fig:PlotsRaw}. The results are shown in figure \ref{fig:InferredEuclidean}. Overlayed are theory curves deduced (via equation \ref{eq:EuclideanDistancePe}) from the Euclidean distances between final states of the ideal GAD. Reiterating theoretical results mentioned above, we see that after approximately two thermalisation times ($t\sim0.1\tau_{sp}$) the coherent state $+X$ outperforms the incoherent ones $\pm Z$. Note that although $-Z$ appears to be \emph{globally} optimal (i.e. has a lower $p_{e}$ at $t\sim0.07\tau_{sp}$ than other states at any time), this is peculiar to our choice of temperatures and is not always the case.

While the behaviour of $p_{e}$ shows good agreement with theory for $\pm Z$ input states, the $+X$ case shows discrepencies in the region $t\leq0.2\tau_{sp}$. This can be ascribed to the fact that our interferometers have finite visibility. This has the effect of mapping some amount of coherence between $\left|H\right>$ and $\left|V\right>$ to an incoherent mixture and is completely analogous to extraneous dephasing processes (e.g. atomic/molecular collisions etc.) that we did not consider in our thermalisation model. Note that $\pm Z$ input states do \emph{not} experience interference effects as they traverse the channels - these states end up traveling through the interferometer via just \emph{one} of the two possible counter-propagating paths and have nothing to interfere with when they re-emerge at the PBS. The $\pm Z$ states are therefore not susceptible to imperfect interferometer visibility. The same cannot be said of the $+X$ state. The interferometers that comprise our channels have typical visibilities $\geq95\%$, though due to the lack of active stabilization, realignment can become necessary from time to time.

\begin{figure}
    \noindent \begin{centering}
    \includegraphics[width=14cm]{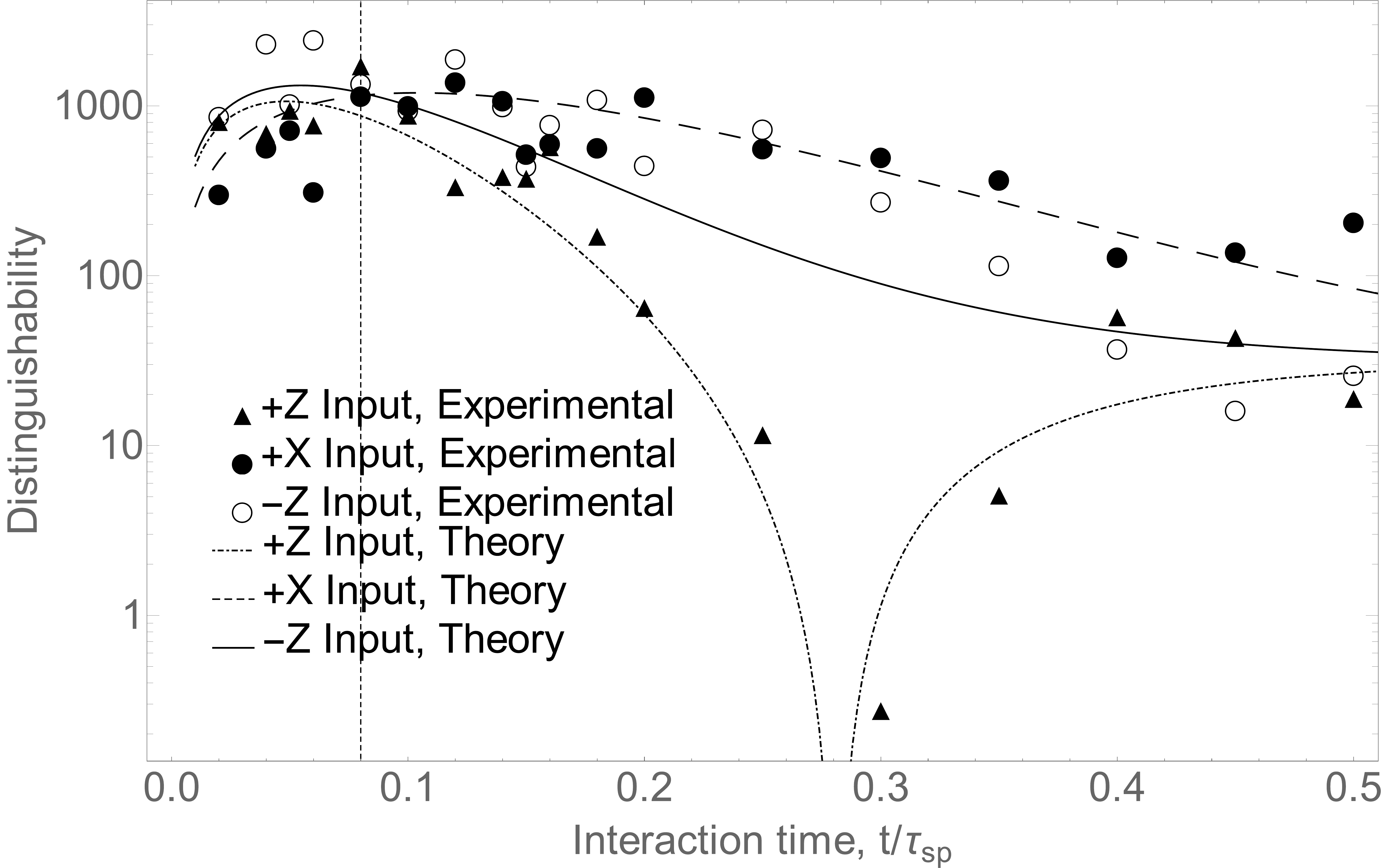}
    \par\end{centering}

    \caption{\textbf{Plot of distinguishability} in the outcome of our optimal
    observable (see figure \ref{fig:PlotsRaw}). Higher points indicate
    better distinguishability. Means and variances were computed over
    multiple sets of comparable experimental data. Average photon number
    in all measurements was $\approx40,000$.\label{fig:Fisher}}
\end{figure}

\begin{figure}
    \noindent \begin{centering}
    \includegraphics[width=18cm]{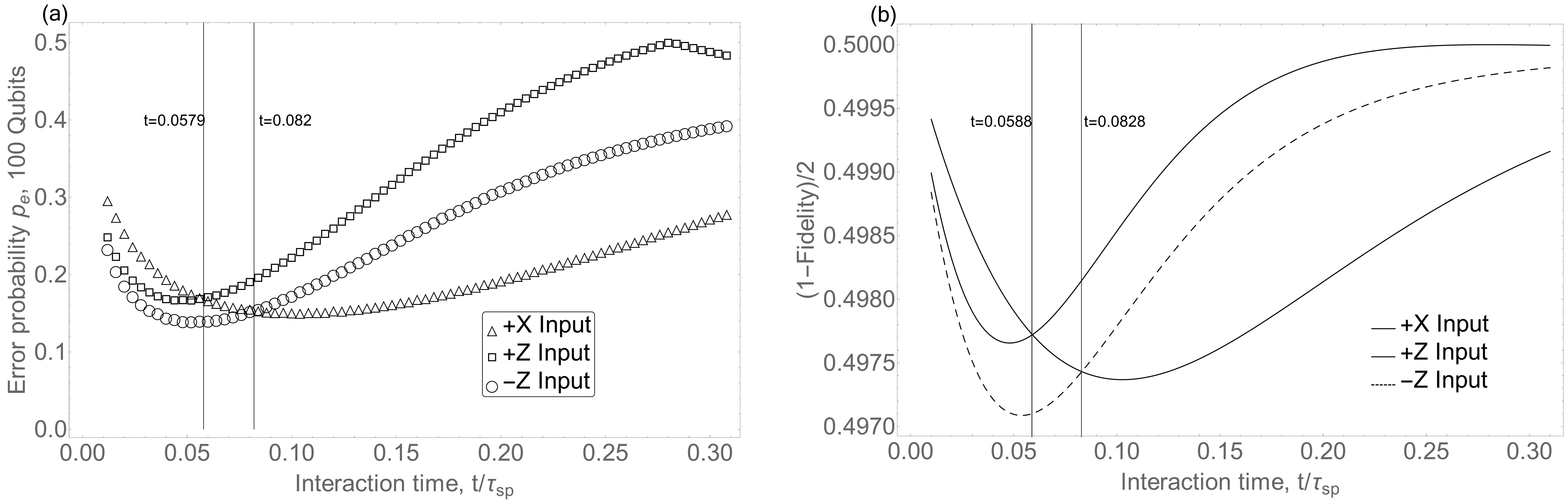}
    \par\end{centering}

    \caption{\textbf{Performance measures for a multi-qubit thermometer. (a)}
    Numerically computed error probabilities for 100 qubits. \textbf{(b)}
    The fidelity between states being discriminated (identical states have
    fidelity 1 whereas orthogonal states have fidelity 0).\label{fig:Multiqubit}}
\end{figure}

While the notion of the Euclidean distance and the error probability are the correct figures of merit to use for state discrimination tasks, the former applies strictly to single-qubit states whereas the latter becomes increasingly difficult to compute for large numbers of qubits. An alternate measure is the distinguishability, often used as a measure of the ease with which two distributions can be distinguished. It is defined as the squared difference of the means of the two distributions, divided by their variance:
\[
\mbox{distinguishability}=\frac{\left(\mbox{difference in mean}\right)^{2}}{\mbox{variance}}=\frac{\left[E\left(P_{hot}\right)-E\left(P_{cold}\right)\right]^{2}}{\mbox{max}\left\{ \mbox{Var}\left(P_{hot}\right),\mbox{Var}\left(P_{cold}\right)\right\} }
\]
where $P_{hot}$ and $P_{cold}$ are the binomially distributed outcomes of some projector observable given the output state from each of the heat baths (i.e. the probabilities shown in figure \ref{fig:PlotsRaw}). A plot of this measure is shown in figure \ref{fig:Fisher}. We attribute the noise in the experimental points in this figure to the fact that relatively few sets of data (9 sets for +Z and +X inputs and 4 sets for -Z) were used to infer the variances. Although qualitatively quite similar to the $p_{e}$ plot in figure \ref{fig:InferredEuclidean} (i.e. the $+X$ state remains optimal after some time, while $-Z$ is optimal if time is not a constraint), the two measures disagree for example on the cutoff times at which the optimal input state changes. A clear advantage for $+X$ is seen at $t>0.08\tau_{sp}$ (from theory curves in figure \ref{fig:Fisher}), whereas in the Euclidean distance or single-qubit $p_{e}$ case it is seen after $t>0.1\tau_{sp}$. This discrepancy leads one to suspect that the optimal input state for our thermometer in a multi-qubit scenario is different from the single-qubit case. In order to obtain a more rigorous measure of many-qubit distinguishability, we numerically computed the error probability $p_{e}$ for 100 qubits. This is shown in figure \ref{fig:Multiqubit} along with the fidelity, $\mathcal{F}=\mbox{Tr}\sqrt{\sqrt{\rho_{hot}}\rho_{cold}\sqrt{\rho_{hot}}}$ where $\rho_{hot}$ and $\rho_{cold}$ are final states from the hot and cold baths respectively. The quantity $\left(1-\mathcal{F}\right)/2$ has been shown\cite{Fuchs1996Thesis} to bound $p_{e}$ from above in the limit of asymptotically large number of qubits. Again, the differences between a single qubit vs many qubits is clear - the crossover between $-Z$ and $+X$ in figure \ref{fig:Multiqubit}(b) occurs at $t=0.0828\tau_{sp}$. In the next section, we treat the many qubit case more carefully.

\section*{Multi-qubit extension and adaptive state discrimination}
\subsection*{A Bayesian approach}

While a single-qubit thermometer is conceptually interesting, it is obviously of more practical relevance to consider a thermometer composed of many spins, but potentially a limited, fixed number. Already, the static strategy - doing the same thing on all $N$ copies of a qubit - yields better binomial statistics the larger $N$ is. Let's return to the example in the previous section, of distinguishing $T=0$ from $T=\infty$. The best static strategy for $N$ qubits in that case is to measure along some optimal axis for all qubits and concluding that the bath is at $T=0$ if and only if \emph{all} measurement outcomes are $0$. In that case, $p_{e}=p_{outcome0}{}^{N+1}$.

One can often do better\cite{mahler2013adaptive,huszar2012adaptive} by allowing for an adaptive strategy (say by changing the measurement basis for each qubit), a possibility that was investigated by Wiseman et al\cite{Wiseman2011StateDisc}. To facilitate further discussion, we now restate the discrimination problem in slightly more formal terms. Whereas in the preceeding discussion we assumed that the bath was equally likely to be in $T_{hot}$ or $T_{cold}$, we now allow each to occur with different prior probabilities (call them $\pi_{1}$ and $\pi_{2}$ where $\pi_{1}+\pi_{2}=1$). We also assume that for a given interaction time $t$, the output states from baths $T_{hot}$ and $T_{cold}$ are known to be $\rho_{1}$ and $\rho_{2}$ respectively. The problem is to find an optimal \emph{strategy} that yields $p_{e,min}=\min_{strat}\left\{ p_{e,strat}\right\} $, the lowest error over all possible strategies. A \emph{strategy}, in turn, is a combination of a measurement (observable $M$) and a \emph{threshold} ($k\in Z$) beyond which one concludes that the output state is $\rho_{1}$ (or $\rho_{2}$). In this language, the error probability for each case (assuming we employ a strategy non-adaptively, i.e. the same $M$ for all $N$ qubits):
\begin{eqnarray}
    p_{e,1} & = & \pi_{1}\sum_{j=0}^{k-1}\frac{N!}{j!\left(N-j\right)!}\left[\mbox{Tr}\left(\rho_{1}M\right)\right]^{j}\left[1-\mbox{Tr}\left(\rho_{1}M\right)\right]^{N-j}+\pi_{2}\sum_{j=k}^{N}\frac{N!}{j!\left(N-j\right)!}\left[\mbox{Tr}\left(\rho_{2}M\right)\right]^{j}\left[1-\mbox{Tr}\left(\rho_{2}M\right)\right]^{N-j}\nonumber \\
    p_{e,2} & = & \pi_{2}\sum_{j=0}^{k-1}\frac{N!}{j!\left(N-j\right)!}\left[\mbox{Tr}\left(\rho_{2}M\right)\right]^{j}\left[1-\mbox{Tr}\left(\rho_{2}M\right)\right]^{N-j}+\pi_{1}\sum_{j=k}^{N}\frac{N!}{j!\left(N-j\right)!}\left[\mbox{Tr}\left(\rho_{1}M\right)\right]^{j}\left[1-\mbox{Tr}\left(\rho_{1}M\right)\right]^{N-j}\label{eq:staticPe}
\end{eqnarray}
The optimal error probability is therefore computed as $p_{e,min}=\min_{k,M}\left\{ p_{e,1},p_{e,2}\right\} $.
If $\rho_{1}$ and $\rho_{2}$ are two single-qubit states, the minimization
over $M$ reduces to a minimization over one real parameter (i.e.
the measurement angle $\theta$; we can assume $\rho_{1}$ and $\rho_{2}$
both lie on the real plane of the Bloch sphere - else we rotate them
onto it).

In the adaptive case where $\theta$ may vary with each copy of $\rho$, the above no longer holds. Instead, consider the following approach: suppose on the first measurement $M_{1}$, we obtain a successful ($\checkmark$) outcome. We update our state of knowledge as follows:
\[
p\left(\rho_{1}|\checkmark\right)=\frac{\pi_{1}p\left(\checkmark|\rho_{1}\right)}{p\left(\checkmark\right)}=\frac{\pi_{1}\mbox{Tr}\left(\rho_{1}M_{1}\right)}{\pi_{1}\mbox{Tr}\left(\rho_{1}M_{1}\right)+\pi_{2}\mbox{Tr}\left(\rho_{2}M_{1}\right)}
\]
\begin{equation}
    p\left(\rho_{2}|\checkmark\right)=\frac{\pi_{2}p\left(\checkmark|\rho_{2}\right)}{p\left(\checkmark\right)}=\frac{\pi_{2}\mbox{Tr}\left(\rho_{2}M_{1}\right)}{\pi_{1}\mbox{Tr}\left(\rho_{1}M_{1}\right)+\pi_{2}\mbox{Tr}\left(\rho_{2}M_{1}\right)}\label{eq:newPriors}
\end{equation}
We then take these as the ``updated'' priors: $\pi_{1}^{\left(1\right)}=p\left(\rho_{1}|\checkmark\right)$ and $\pi_{2}^{\left(1\right)}=p\left(\rho_{2}|\checkmark\right)$ where the superscript indicates that they are post-measurement-1. Had the outcome of $M_{1}$ been negative, we would simply have replaced every instance of $\mbox{Tr}\left(\rho M\right)$ in equation \ref{eq:newPriors} with $1-\mbox{Tr}\left(\rho M\right)$. We then return to equation \ref{eq:staticPe} to minimize $\theta$ substituting $\pi_{1}\to\pi_{1}^{\left(1\right)}$ and $\pi_{2}\to\pi_{2}^{\left(1\right)}$. Now armed with $M_{2}$, we proceed to repeat the procedure on the next copy of $\rho$. After all qubits are measured, we finally choose the state associated with the larger of $\left\{ \pi_{1}^{\left(n\right)},\pi_{2}^{\left(n\right)}\right\} $ as the output of our discrimination procedure. In \cite{Wiseman2011StateDisc}, it was shown that such a Bayesian prior update strategy is not only better than the static one but is in fact optimal if $\rho_{1}$ and $\rho_{2}$ are pure states! And although it isn't the optimal strategy for an arbitrary mixed state, is nevertheless performs better than the static approach.

\subsection*{Testing Adaptivity}

To test the adaptive approach, we set the channel to identity (it simply preserves all input states; in practice this meant setting $p=0$ in the VBS and $\theta_{H}=\theta_{V}=0\text{\textdegree}$ for the channels). The states to discriminate were $\left|H\right>$ and $\left|D\right>$. These states were chosen (instead of outputs of the thermalisation channel described in preceeding sections) because the identity channel is much easier to control, allowing us to avoid potential confounding factors when demonstrating the benefits of adaptivity. The identity channel requires only one of channels 1 and 2 to be active and so does \emph{not} require LCWP calibration or careful alignment of the VBS, among other things.

\begin{figure}
    \noindent \begin{centering}
    \includegraphics[width=14cm]{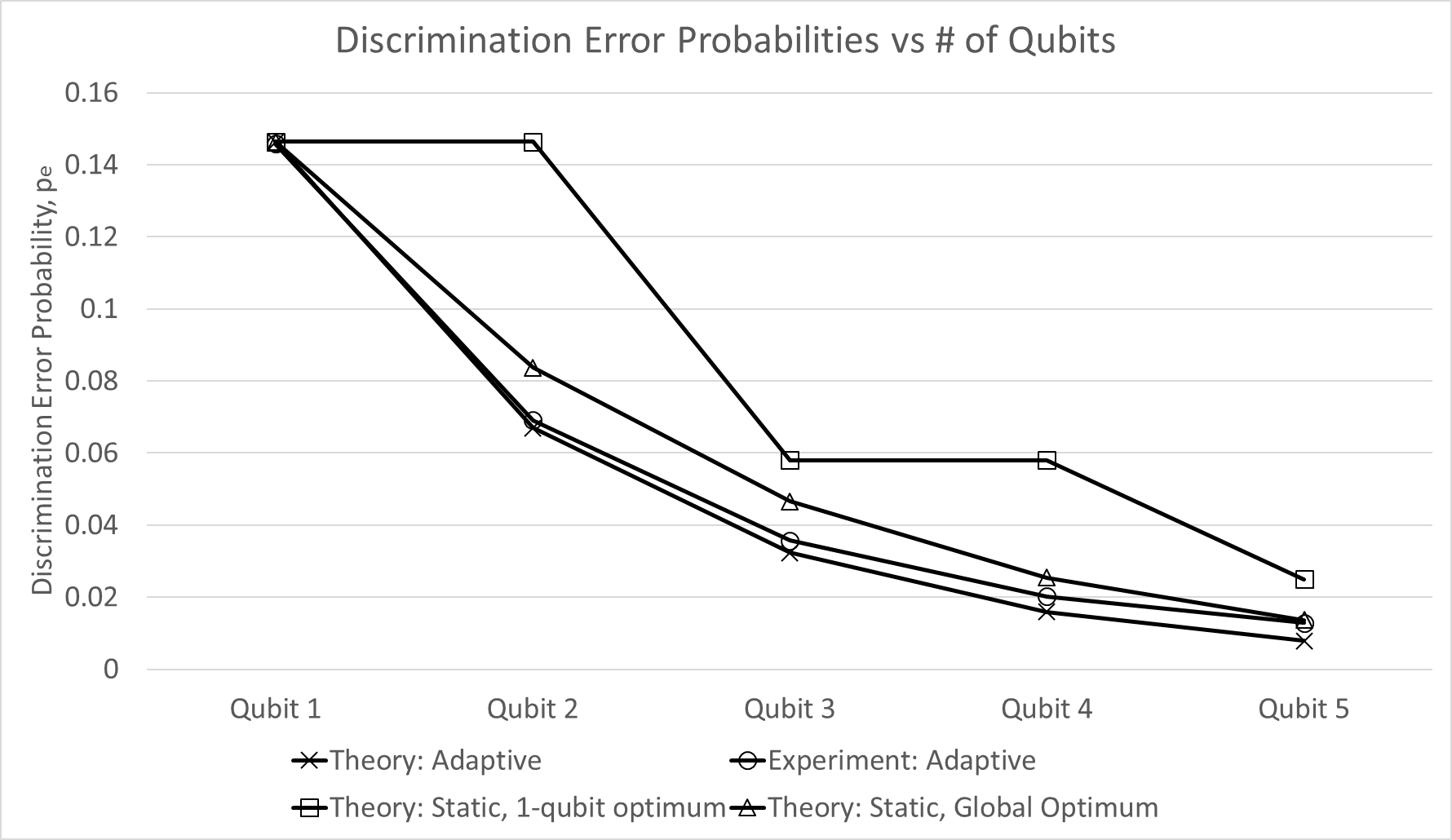}
    \par\end{centering}

    \caption{\textbf{Plot of error probability} in discriminating $\left|H\right>$
    from $\left|D\right>$, by number of qubits for various strategies.\label{fig:AdaptivePe}}
\end{figure}

As before, we began by validating the channel via process tomography and then proceeded to measure along bases specified by equations \ref{eq:staticPe} and \ref{eq:newPriors} above. In an actual adaptive scenario, one ideally updates one's measurement setting as each photon is detected. However, our motorised waveplate mounts are relatively slow making it difficult for us to adapt our measurement setting conditioned on individual detection events. Instead, we have chosen to measure along \emph{all} bases that are prescribed by our Bayesian update strategy given \emph{all }possible detection outcomes. This allowed us to map out the full tree of possible outcomes along with the probability of occurrence for each node. Such a map of outcome probabilities allows us to confirm that the Bayesian update rule is valid, even if we can't directly emulate adaptivity. Figure \ref{fig:AdaptivePe} shows a plot of how the error probability, $p_{e}$, scales with number of qubits in the various scenarios.

Figure \ref{fig:AdaptivePe} shows the resulting error probabilities, $p_{e}$, for various strategies. Theoretical predictions and experimentally derived values for our strategy, detailed above, are plotted in blue and red, and are labelled ``Adaptive''. For comparison, theory predictions for two static (non-adaptive) cases are shown. The ``1-qubit optimum'' naively uses a measurement basis that is optimal for just 1 qubit, and repeats it as necessary. On the other hand, in the ``Global Optimum'' approach, one is assumed to have been told the total number of qubits available, and a \emph{static} measurement angle that is optimal for the given number of qubits is computed and used.

Evidently, the \emph{adaptive }multi-qubit scenario offers the benefit of lower error probabilities. The absolute reduction is particularly pronounced for the first few additional qubits, when $p_{e}$ is still relatively large. The deviation between our data and theoretical prediction for $p_{e}$ becomes pronounced as the number of qubits $N$ becomes large. The probability tree that we must reconstruct grows quickly with $N$ (generally, $2^{N}$ branches) and so does the precision with which we must set our measurement basis in order to maintain an advantage over the non-adaptive approach. In our case, we are limited by our motorised waveplate mounts to an angular precision no better than $\pm0.2\text{\textdegree}$.

Also of note, our adaptive strategy makes no assuption about the total number of qubits and continues to work in a ``rolling'' fashion even if, midway through the scheme, we were suddenly told that more qubits have suddenly become available. This is in stark contrast to the ``global optimum'' static strategy, where a favorable scaling is only possible given full knowledge of just how many qubits there are.

\section*{Conclusions}

In summary, we have simulated the thermal equilibration of a spin by using a construction of a universal emulator for quantum channels. This has allowed us to confirm Jevtic et al's\cite{rudolphThermometry2015} theoretical conclusions that for thermalisation with a bosonic bath, optimal temperature discrimination occurs at early times rather than in the asymptotic limit, and that for most interaction times, a thermometer initialized in a coherent superposition state outperforms one prepared in the ground state. In our case, this advantage translates to a maximum reduction of the error probability, $p_{e}$, from 47.99\% to 46.19\% when discriminating between temperatures $5.98\hbar\omega/k_{B}$ and $10\hbar\omega/k_{B}$ after letting the qubit interact with the bath for $t=0.23\tau_{sp}$. This is a 90\% or almost two-fold increase in the improvement over a purely random guess ($p_{e}=50\%$). Furthermore, after just $t=0.12\tau_{sp}$, the error probability is reduced, relative to a fully thermalised qubit, from 49.17\% to 46.12\% or approximately a 3.5 fold advantage in the improvement over random guess. We discuss the origin and limitations of this behaviour, and study the extension to the case of a thermometer composed of a finite number of spins, showing the advantages of an adaptive measurement strategy. We note that there are important differences between the optimization problem for single and multiple spins, but conclude that quantum coherence retains an advantage even in the latter case. This is a new example of a quantum metrological advantage, and may prove important for making accurate measurements of thermal properties of quantum systems with limited resources or limited disturbance.

\textit{Note:} During preparation of this manuscript we became aware that similar work was being pursued by Mancino et al\cite{Marco}.

\bibliography{ChannelsThermometry}

\section*{Acknowledgements}

We thank D. Mahler and B. Sanders for useful discussions. This work was funded by NSERC, CIFAR, and Northrop-Grumman Aerospace Systems \textit{NG Next}.

\section*{Author contributions statement}

A.M.S., A.V.S., H.F., and W.K.T. jointly designed the experiment. W.K.T. performed the experiment and corresponding data analysis. A.M.S., H.F., and W.K.T. jointly conducted theoretical analysis of the qubit thermometry scheme and Bayesian adaptive state discrimination. W.K.T. wrote this manuscript along with contributions from A.M.S. and H.F.

\section*{Additional information}

\textbf{Competing financial interests:} The authors declare no competing
financial interests.

\end{document}